\documentclass[twocolumn]{aastex61}

\newcommand\aastex{AAS\TeX}

\submitjournal{ApJL}

\shorttitle{\aastex\ 1I/`Oumuamua As a Tidal Disruption Fragment}
\shortauthors{\'Cuk}

\begin{document}

\title{1I/`Oumuamua as a Tidal Disruption Fragment From a Binary Star System}

\correspondingauthor{Matija \'Cuk}
\email{mcuk@seti.org}

\author{Matija \'Cuk}
\affil{SETI Institute \\
189 N. Bernardo Ave, Suite 200 \\
Mountain View, CA 94043, USA}

\begin{abstract}

1I/`Oumuamua is the first known interstellar small body \citep{bac17}, probably being only about 100~m in size. Against expectations based on comets, `Oumuamua does not show any activity and has a very elongated figure \citep{mee17}, and also exhibits undamped rotational tumbling \citep{fra17}. In contrast, `Oumuamua's trajectory indicates that it was moving with the local stars, as expected from a low-velocity ejection from a relatively nearby system \citep{mam17}. Here I assume that `Oumuamua is typical of 100-m interstellar objects, and speculate on its origins. I find that giant planets are relatively inefficient at ejecting small bodies from inner solar systems of main-sequence stars, and that binary systems offer a much better opportunity for ejections of non-volatile bodies. I also conclude that `Oumuamua is not a member of a collisional population, which could explain its dramatic difference from small asteroids. I observe that 100-m small bodies are expected to carry little mass in realistic collisional populations, and that occasional events when whole planets are disrupted in catastrophic encounters may dominate interstellar population of 100-m fragments. Unlike the Sun or Jupiter, red dwarf stars are very dense and are capable of thoroughly tidally disrupting terrestrial planets. I conclude that the origin of `Oumuamua as a fragment from a planet that was tidally disrupted and then ejected by a dense member of a binary system could explain its peculiarities. 

\end{abstract}


\section{Introduction} \label{sec:intro}

1I/`Oumuamua is the first known interstellar object, that was discovered in October 2017 after it has already passed perihelion \citep{bac17, mee17}. `Oumuamua is clearly extrasolar in origin, having a velocity at infinity of 26 km/s and an eccentricity of 1.2. Much of Oumuamua's velocity relative to the Solar System is a reflection of the Sun's own velocity relative to the local standard of rest \citep{mam17}, making `Oumuamua's trajectory close to our expectations for an interstellar interloper. However, `Oumuamua's physical characteristics were unexpected, starting with a complete lack of cometary activity \citep{kni17, mee17} or associated meteoroids \citep{ye17}. The observed spectrum of `Oumuamua is featureless and somewhat red \citep{mas17, ban17}, similar to a number of outer Solar System objects. However, `Oumuamua's Solar System spectral analogues are expected to be volatile rich and should exhibit cometary activity after passing within 0.25 AU from the Sun, as `Oumuamua did. Therefore, one cannot say if `Oumuamua's spectral similarity to certain outer Solar System bodies is meaningful or a coincidence. The most puzzling feature of `Oumuamua is its very elongated shape, with aspect ratio of 5:1 to 10:1 \citep{bol17, mee17}. This is an extreme value for Solar System bodies of similar size, and may indicate that `Oumuamua has internal strength \citep{fra17}. While `Oumuamua's rotation period was reported to be 7-8 h, it has been suggested that the observations are not consistent with a single period, probably indicating a non-principal axis rotation, i.e. tumbling \citep{fra17, dra17}. Non-damped tumbling would indicate that the interior of `Oumuamua is not particularly dissipative, and is consistent with `Oumuamua being a rigid body \citep{fra17}. Monolithic 100-m bodies are known in the Solar System, but are less common than ``rubble piles", tend to have less elongated shapes and, not being at risk from rotational disruption, often have very short spin periods \citep[as radiational YORP effect distributes their rotation rates throughout the large allowed phase space;][]{pra02}.  

Therefore, while the trajectory and rotational physics of `Oumuamua appear to be consistent with our prior understanding, its shape and composition are surprising. In this paper, I will make an assumption of Copernican principle with respect to `Oumuamua, i.e. that it is typical of bodies of its size that are populating the local interstellar space. This assumption is far from secure as it is based on only one object, but it is also testable as we expect to detect more interstellar bodies as more and larger automated surveys become operational.

\section{Dynamical Considerations} \label{sec:dynamics}

Ejection of planetesimals from our Solar System is a natural consequence of planetary formation and migration \citep{fer84, dun87, kai08}. Young giant planets scattered the remaining small bodies, with the resulting exchange of angular momentum enabling expansion of the orbits of Neptune, Uranus and Saturn, which on average were passing small bodies from the trans-Neptunian belt to Jupiter. Jupiter, due to its large mass, was highly efficient at ejecting bodies from the system, resulting in the planet's inward migration. Some of the bodies that narrowly escaped ejection ended up on very large orbits torqued by passing stars and Galactic tide, forming the Oort Cloud. The existence of the Oort Cloud, inferred from the continuous influx of long-period comets, is a direct indication that large number of comets must have been ejected from our Solar System when the Oort Cloud formed.

Ejection of volatile-free asteroids is also possible, but they are thought to have been a relatively small fraction of planetesimals that were ejected or placed into the Oort cloud. One reason for this is the much greater supply of icy planetesimals in our system, which may not apply elsewhere. Another is that our giants planets all orbit beyond the ``snowline", the distance beyond which planetesimals incorporate large fractions of water ice. Rocky planetesimals are thus less likely to make it to Jupiter-crossing orbits, and the majority of bodies with $a<$2.5 AU that become unstable end up colliding with the Sun \citep{gla97, min10}. 

Many known giant exoplanets are well within the relevant ``snowline" for their solar system, and some of them are certainly capable of ejecting small bodies into interstellar space \citep{for08}. In general, two major processes are competing for the elimination of planetesimals: scattering and collisions. Collisions become more likely if the planet is brought closer to the star (as size of the planet increases relative to the size of the orbit), but scattering becomes less efficient closer to the star, as the orbital velocities are higher while the the planet's escape velocity is the same. Using expressions from \citet{tre93} and \citet{wya17}, I find that the ratio of ejection and collision rates is (assuming constant density for the planet):
\begin{equation}
{R_{ej} \over R_{col}} \propto {M_p^{4/3} a^2 \over M_{*}^2}
\label{ratio}
\end{equation}
Where $M_p$ and $M_*$ are masses of the star and the planet, and $a$ is the planet's semimajor axis. This implies that Jupiter at 1~AU would be more than an order of magnitude weaker ejector than at its present distance. Since for most of the Main Sequence, $\log(L_*/L_{\rm Sun})=k \log(M_*/M_{\rm Sun})$, with $3.5<k<4$ \citep{dur04}, it is clear from Eq. \ref{ratio} that Jupiter mass planets at Habitable Zone distances from their parent stars ($(a/1~\rm{AU})^2 = L_*/L_{\rm Sun}$) would be less efficient scatterers for smaller stars, but more efficient for more massive stars. Also, planets more massive than Jupiter could be efficient at ejection rocky planetesimals even around solar-mass and smaller stars. 

However, here I am making the assumption that `Oumuamua is typical among interstellar objects, which implies that volatile-free ejected bodies are more common than comets. Even in situations in which planets can efficiently eject rocky planetesimals (massive planets, high-mass stars), it appears unlikely that they could dominate the galactic population of scattered bodies. Even relatively low-mass planets can eject comets at large heliocentric distances (Neptune being an example), while low-mass main-sequence stars have an order of magnitude closer-in snowlines, implying less space available for purely rocky planetesimals. Additionally, large mass available in volatiles should in general make the mass of solids beyond the snowline larger than that available in the inner system, where only refractories are stable. Therefore I come to the conclusion that, on the basis of our present knowledge, we would expect most interstellar small bodies be cometary, i.e. to be formed beyond the snowline of their solar systems and to contain significant amounts of volatiles. Lack of any cometary activity from `Oumuamua is therefore unexpected and should make us rethink about dominant mechanisms for formation of 100~m fragments on interstellar trajectories. This is independent of any issues related to the shape of `Oumuamua, which will be discussed in the next section.
 
At this point, I note that binary and multiple stars are very common in the Galaxy, and that they are likely to be major source of ejected orbits \citep{smu16, wya17}. Binary star and planetary populations are quite distinct, as they are separated by the "brown dwarf desert", at least as companions to main sequence stars. Therefore the populations of bodies ejected by binary companions and planets may turn out to be quite distinct. Binary separations cover a wide range of values, with some companions being closer than 1~AU \citep{duc13}. Of course, availability of material for ejection is still an issue for binary companions, and one may still expect icy material to predominate among small bodies in binary systems. But, if `Oumuamua is an indication that the events leading to ejections of most numerous interstellar bodies are very different than those that operate in our Solar System, it is a very good guess that stellar companions may be implicated, as they are both very common, and more efficient scatterers than planets.\footnote{After the first version of this Letter was submitted for review, I realized that \citet{ray17} reached conclusions similar to mine on the relative abundance of icy and rocky planetesimals, as well as on the possible importance of binary stars for ejection of small bodies}. Interestingly, most likely candidates for the origin of `Oumuamua identified by \citet{zul17} are binary systems, but the probability of any individual system being the ultimate source of `Oumuamua is rather small. 

\section{Physical Characteristics} \label{sec:physics}

Despite passing within 0.25 AU from the Sun, `Oumuamua did not exhibit any measurable cometary activity \citep{kni17, mee17}. Since `Oumuamua's visible spectrum is similar to some moderately red outer Solar System bodies, there have been suggestions \citep{lau17} that the interior of `Oumuamua is volatile rich, but that the surface has been devolatilized by very prolonged exposure to conditions of interstellar space. However, this hypothesis is in conflict with the fact that Oort Cloud comets are experiencing an environment practically indistinguishable from interstellar space for billions of years, yet they exhibit cometary activity when approaching the Sun. The most likely explanation for `Oumuamua's lack of outgassing is that it is inherently volatile-poor. 

Comparison of `Oumuamua to asteroids of similar size also raises questions. The spectrum of `Oumuamua, being both moderately red and featureless, is not common in the inner Solar System \citep{ban17}. Also, `Oumuamua's extreme elongation, between 5:1 \citep{fra17} and 10:1 \citep{mee17} makes it an outlier among asteroids \citep{pra02}. `Oumuamua's undamped tumbling indicates that it may be a monolith, rather than a ``rubble pile" \citep{fra17}. Asteroidal monoliths in the 100-meter range are known in the Solar System, and considering a relatively long (for a monolith) 7-8 hour rotation period, undamped rotation by itself is not remarkable \citep[for larger asteroids tumbling is usually reserved for even slower-spinning bodies;][]{pra02, bur73, sha05} and does not need any special excitation mechanism.  

In the Solar System, 100~m bodies are thought to be collisional fragments of larger progenitors. Current theories of planetesimal formation indicate that original planetesimals may have been much larger, in the 100~km range \citep{you05, joh07, mor09}. Asteroids (433) Eros (30~km long) and 25143 Itokawa (600~m long), both visited by spacecraft, are thought to be typical of intermediate steps of collisional evolution from 100~km bodies to 100~m fragments. Eros has a density of 2.7 g/cc, similar to ordinary chondrite meteorites it is likely related to, and is thought to be a fractured body, meaning that it is held together by gravity but without significant internal voids \citep{che04}. Itokawa, in contrast, has a density of 1.9 g/cc despite a composition similar to Eros's. Itokawa is thought to be a rubble pile, with fragments of a range of sizes held together by gravity and possibly E-M surface forces, with a large porosity \citep{fuj06}. While solid blocks are present on Itokawa, they are results of many collisional events, some of which destroyed their past parent bodies, and some led only to fracturing. It is easy to see why long and thin fragments would be rare, as the orientation of stresses from multiple asteroidal collisions becomes basically random. Therefore, not only is the elongation of Oumuamua unusual (near-Earth asteroid 1865 Cerberus comes close to this, but it is over 1~km long and probably not a monolith), but one would expect such long and thin pieces to be rare on theoretical grounds. While `Oumuamua could be an outlier from a population similar to asteroids, here I am assuming that `Oumuamua is typical of interstellar objects, which would make them collectively quite distinct from asteroids. 

Interstellar nature of `Oumuamua has led to speculation of a possible artificial nature. However, its trajectory is that of ``celestial driftwood"\footnote{Michele Bannister, 11/21/17, Twitter} as shown by \citet{mam17}. If `Oumuamua were to be artificial, it would require artificial 100-meter bodies on passive interstellar trajectories to be more common than ejected asteroids and comets of the same size. This is an extraordinary claim, and would require evidence more extraordinary than `Oumuamua's elongated shape. More specifically, continuing tumbling of `Oumuamua is most consistent with a single solid body with no moving parts \citep{fra17}. A hollow object containing movable items would damp its non-principal axis rotation much more quickly \citep{bur73}. Therefore, an artificial origin would not explain any of `Oumuamua observed peculiarities.

If `Oumuamua is not a result of collisional evolution like the one experienced by Solar System asteroids, is there a way of naturally producing its shape? Collisions between the original 100-1000 km planetesimals (or planets accreted from them) would not be pre-fractured or reaccreted, so they could produce a rather different suite of first-generation collisional fragments, some of which may have unusual shapes. But the amount of mass in 100-meter fragments resulting from disruption of much larger bodies would be modest (as most of the mass would be in larger pieces). Also, these unprocessed planetesimals would have to be on somewhat stable orbits in order to have a reasonable probability of colliding, but then their fragments would need to be ejected by a giant planet or a companion star rapidly, before a collisional cascade could obliterate non-compact shapes. This scenario is rather contradictory and unlikely to produce more interstellar bodies than ejection of more conventional collisional fragments.

Interestingly, `Oumuamua is at or just below the size threshold at which asteroid material strength starts dominating over gravity \citep{obr03}. It is tempting to speculate that if a larger body's gravity would somehow be neutralized, it may dissociate itself into fragments the size of `Oumuamua (if other conditions are met). Gravity, of course, cannot be ``turned off", but gravitationally bound bodies can be torn apart by variety of mechanisms, a prominent one being tidal forces. Comet Shoemaker-Levy 9 was dramatically tidally disrupted by Jupiter, before its collision with the planet \citep{asp96}. It is fully possible that some of the comets that were ejected by Jupiter may have first been tidally disrupted. However, tidal disruption in the Solar System is limited by the restricted density ranges of planets and small bodies. Jupiter's density of 1.3 g/cc means that a comet must approach very closely (within a couple radii) to the planet in order to be disrupted, limiting the efficiency of the process \citep{jef47, hol08}. Also, most tidal disruptions of rubble piles are marginal, with the pieces reaccreting into one or more new ruble piles afterward \citep{wal06}. 

In the previous section I proposed that stellar companions are expected to be major producers of interstellar small bodies in the Galaxy. A solar type star has a density comparable to Jupiter's, so tidal disruptions are unlikely to be very common during ejections. However, M-type main sequence stars are significantly denser than the Sun, with M0V stars being 3 times, M5V 20 times, and M9V stars 150 times more dense than the Sun \citep{kal09}. Such densities make it possible for these low mass stars to tidally disrupt not only under-dense rubble piles, but also planet-sized bodies. An Earth-like planet passing close enough to an M-type dwarf would be completely torn apart, with only material cohesion on sizes comparable to that of `Oumuamua being able to resist the star's tides. In addition, sizable planets would also suffer decompression during tidal encounters, which would help disperse the fragments during this event \citep{asp06}. Interior material would be exposed to vacuum, which could lead to rapid solidification and other strange effects. Even if the planet was not volatile poor, atmosphere and volatile-rich layers may not survive the event, both due to tidal forces and stellar irradiation (however brief). Anisotropic forces acting to shape fragments may produce elongated bodies like `Oumuamua, and with moderate rotation rates as observed for `Oumuamua, in contrast to monoliths produced in collisions. 

\section{Numerical Test}\label{sec:numerical}

After the original version of this paper was submitted (and a preprint made public), \citet{jac17} have published numerical simulations of planetesimal scattering in binary systems. \citet{jac17} state that their simulations did not find any cases where a planet would be tidally disrupted by a close stellar passage before being ejected, indicating that such disruptions are very rare. In this section I will address the claims of \citet{jac17} with the help of a simple numerical simulation.

\begin{figure*}
\plotone{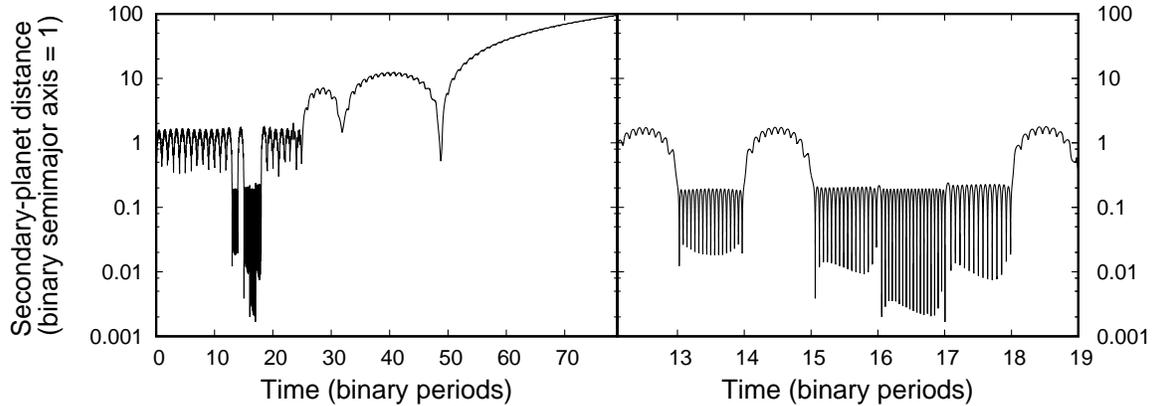}
\caption{Simulation of a binary star system with a planet initially orbiting the primary, done using {\sc rebound}'s IAS15 algorithm. The planet initially orbits the larger star, but then gets temporarily captured into orbiting around the smaller star during binary orbits 13-14 and 15-18(enlarged in right-hand panel). The minimum separation between the planet from the smaller star is $1.7 \times 10^{-3}$ of the binary semimajor axis. Around the fiftieth binary period the planet is ejected from the system.\label{inner1}}
\end{figure*}

Figure \ref{inner1} shows a numerical simulation of a binary system, with a planet initially orbiting the more massive component. Stellar mass ratio is 2:1, the planet is treated as massless, the ratio of binary and planetary semimajor axis is 6:1, both eccentricities are 0.5, while the planet's orbit is inclined by $10^{\circ}$. Initially, both the planet and secondary are at their periastra (which are aligned), and the planet is also at its ascending node. The simulation was done using the IAS15 algorithm \citep{rei15} within the {\sc rebound} integration package \citep{rei12}. Figure \ref{inner1} plots the distance between the planet and the smaller component. At first, the planet orbits the larger star (for most of $0 < t < 25$),  occasionally it orbits the smaller star ($13<t<14$ and $15<t<18$), then the planet orbits both stars on a large orbit ($25< t < 50$), and finally the planet is unbound ($t > 50$). 

While the planet is orbiting the less massive star, mutual distance is at one point only $1.7 \times 10^{-3}$ of the binary's semimajor axis. A similar ``temporary capture" has happened to comet Shoemaker-Levy 9 before its tidal disruption and impact on Jupiter \citep{kar96}, and was likely important during capture of Jupiter's irregular satellites \citep{cuk04}. Temporary capture typically begins and ends when the stars are at periastron, so the less massive star's Hill sphere is at its smallest. Figure \ref{inner1} shows that the most eccentric temporary capture orbits (and the closest approaches to the smaller star) occur during entrance into and exit from temporary capture.

Our simulation results are unit-independent and can be applied to a range of binary masses and separations. If the binary semimajor axis is set to 1~AU, then the closest approach of the planet to the secondary would be $2.5 \times 10^{5}$~km, or 0.36~$R_{\rm Sun}$. This distance results in a collision for main sequence stars of the spectral type M4 (with $\simeq 0.5 M_{\rm Sun}$) and earlier. For later type M dwarfs the collision is avoided, but the tides would totally destroy any plausible planet. The least massive main sequence star, a $0.075~M_{\rm Sun}$ M9 dwarf, would be able to tidally disrupt a fluid non-rotating planet with a density of 30 g/cc \citep{har96}, and the tides from M5V-M8V stars would be even stronger.

Simulation shown in Fig. \ref{inner1} was literally the first simulation I attempted of an instability with a planet starting on a S-type orbit \citep[orbiting one of the stars; nomenclature from][]{hol99}. \citet{jac17} in contrast integrated only P-type orbits, where their particles initially orbited both stars. I also did a quick test of planets initially on P-type orbits, and obtained results consistent with those of \citet{jac17}, as the planet never comes very close to either of the stars before being ejected. Therefore, results of \citet{jac17} show that tidal disruption is unlikely only for initially P-type planetary orbits, but have no relevance for S-type orbits, which appear to hold much more potential for tidal disruption. 

Note that the planet (or several, since an instability is required to couple the planet to the binary companion) would need to form within 20\% of binary periastron distance \citep{qui07}, or 0.1 of binary semimajor axis in this case. Almost half of red dwarfs are thought to have super-Earth's in their habitable zones \citep{bon13}, and the habitable zone of 0.1~AU corresponds to a M4V star \citep{kal09}, so a binary separation of 1 AU is not in conflict with planets forming close to the primary. For larger stellar separation, optimal size of disrupting star would move to somewhat larger masses.

\section{Discussion and Conclusions}\label{sec:discussion}

Here I are proposing that `Oumuamua is a part of dominant population of 100-meter interstellar objects that were generated in tidal disruptions of solid planets by M-dwarfs in binary systems. Many of the resulting fragments should be of size when material forces become more important than gravity, that is hundreds of meters. Fragments would generally be volatile poor, and their shapes may be quite irregular, possibly elongated. Unless the original planet was a bound companion of the red dwarf, the fragments would almost certainly be ejected from the system, likely without any significant collisional evolution.

This is a somewhat exotic way of producing small bodies, quite different from our experience based on the Solar System. However, `Oumuamua is clearly suggesting that the range of processes operating on 100-m bodies in the Galaxy extends beyond those we are familiar with. First of all, it is very likely that binary stars are a very important contributor to the population of interstellar asteroids, as stars are naturally more powerful scatterers than the planets. Second, if occasional extreme events are able to produce large numbers of 100-m fragments, such bodies may overwhelm the population of collisionally-produced comets and asteroids that are ejected individually by either planets of binary companions. 

How common are objects like `Oumuamua? Published estimates of their number density based on one detection include $10^{15}$ pc$^{-3}$ \citep{por17, mee17} and $10^{16}$ pc$^{-3}$ \citep{tri17}. Since there is no more than one star per cubic parsec in the Galactic disk \citep{por17}, this would mean that there are at least $10^{15}$ `Oumuamuas for each star in the disk. If we assume an albedo of 0.2 and therefore dimensions of $180 \times 18 \times 18$ meters, equivalent volume sphere radius is about 40 meters (5:1 aspect ratio does not change that). A total of $10^{15}$ such objects, with a density of 3 g/cc would have a mass comparable to that of Mars ($0.1 M_{\rm Earth}$). On the other hand, if the Solar System ejected approximately $10~M_{\rm Earth}$ of 100~km icy planetesimals \citep{nes16}, and the overall cumulative size distribution of TNOs below 100~km is proportional to $D^{-2}$ \citep{bie15, gre15, rob17}, then I estimate that approximately $0.01~M_{\rm Earth}$ of 100-m bodies was ejected by the planets from our Solar System. Therefore, if the Sun is typical, mass in `Oumuamua-like objects may be ten times larger than in interstellar comets of the same size (the number ratio is a factor of few closer due to lower cometary densities). Actually, the average stellar rate of cometary ejections may be below solar, as comets are easier to detect than asteroids like `Oumuamua \citep{eng17, mee17}. 

A Mars-mass of `Oumuamuas ejected from every system seems rather excessive, especially given that this is the low end of available estimates of the number of `Oumuamuas. Not every star is a binary, and not all binaries include a dense M-dwarf. Therefore, I speculate that a smaller number of larger events generates the observed population. So the kind of ``typical" disruption I am envisioning is that in every hundredth solar system, a $10 M_{\rm Earth}$ super-Earth has a close encounter with a dense late M-dwarf and is tidally disrupted into 100-meter strength-dominated fragments with a relatively high efficiency, and that these fragments are then ejected from the system. M-dwarfs are the most common stars, while we have recently learned that possibly as many as half of the stars have super-Earths \citep{buc12}, so an occasional event of this type would not be extraordinary. 

The present  hypothesis is based on the assumption that `Oumuamua is not a fluke but a typical representative of interstellar asteroids. If future discoveries look more like Solar System comets or small asteroids, the need for this exotic formation mechanism becomes less pressing. But if new discoveries are likewise monolithic fragments with unusual shapes, defying the expectations based on collisional evolution, this idea may warrant a closer look, with in-depth modeling of binary system dynamics and tidal disruptions being needed before we can determine if this hypothesis is tenable.   
 
\acknowledgments

M\'C acknowledges Igor Smoli\'c as the first to suggest the possibility that `Oumuamua is tumbling, which inspired the current paper. The author thanks an anonymous referee for suggestions that greatly improved the manuscript, as well as Michele Bannister, Igor Smoli\'c, Dan Tamayo and Jorge Zuluaga for their helpful comments on the first version of the paper. Simulations in this paper made use of the {\sc rebound} code which can be downloaded freely at http://github.com/hannorein/rebound. M\'C is supported by NASA Emerging Worlds award NNX15AH65G.

%% This command is needed to show the entire author+affilation list when
%% the collaboration and author truncation commands are used.  It has to
%% go at the end of the manuscript.
%\allauthors

%% Include this line if you are using the \added, \replaced, \deleted
%% commands to see a summary list of all changes at the end of the article.
%\listofchanges

\end{document}